%% file: 0_main.tex
\documentclass[conference]{IEEEtran}
\IEEEoverridecommandlockouts
\usepackage{cite}
\usepackage{amsmath,amssymb,amsfonts}
\usepackage{algorithmic}
\usepackage{graphicx}
\usepackage{textcomp}
\usepackage{xcolor}
\usepackage[colorlinks=true,linkcolor=black,anchorcolor=black,citecolor=black,filecolor=black,menucolor=black,runcolor=black,urlcolor=black]{hyperref}
\usepackage{siunitx}
\usepackage{booktabs}

\usepackage[mathscr]{euscript}

\usepackage{stfloats}
\def\BibTeX{{\rm B\kern-.05em{\sc i\kern-.025em b}\kern-.08em
    T\kern-.1667em\lower.7ex\hbox{E}\kern-.125emX}}

\usepackage{comment}
\usepackage{tabularx}
\usepackage{longtable} % tables that can span several pages
\usepackage{colortbl}
\usepackage{multirow, multicol, array, makecell}
\usepackage{cleveref}
\usepackage{subcaption}
\crefname{section}{Section}{Sections}
\crefname{figure}{Figure}{Figures}
\crefname{table}{Table}{Tables}

\usepackage{glossaries}
\newacronym{cnn}{CNN}{Convolutional Neural Network}
\newacronym{dl}{DL}{Deep Learning}
\newacronym{vc}{VC}{Voice Conversion}
\newacronym{tts}{TTS}{Text-to-Speech}
\newacronym{mfcc}{MFCC}{Mel Frequency Cepstral Coefficient}
\newacronym{stft}{STFT}{Short Time Fourier Transform}
\newacronym{cqcc}{CQCC}{Constant Q Cepstral Coefficient}
\newacronym{roc}{ROC}{Receiver Operating Characteristic}
\newacronym{auc}{AUC}{Area Under the Curve}
\newacronym{tp}{TP}{True Positive}
\newacronym{tn}{TN}{True Negative}
\newacronym{dft}{DFT}{Discrete Fourier Transform}
\newacronym{mae}{MAE}{Mean Absolute Error}
\newacronym{svm}{SVM}{Support Vector Machines}
\newacronym{gmm}{GMM}{Gaussian Mixture Models}
\newacronym{gnn}{GNN}{Graph Neural Network}
\newacronym{mlp}{MLP}{Multi-Layer Perceptron}
\newacronym{ssl}{SSL}{Self-supervised Learning}
\newacronym{asv}{ASV}{Automatic Speaker Verification}
\newacronym{eer}{EER}{Equal Error Rate}
\newacronym{ir}{IR}{Impulse Response}
\newacronym{snr}{SNR}{Signal-to-Noise Ratio}
\newacronym{oc}{OC}{One-Class}
\newacronym{stfpm}{STFPM}{Student-Teacher Feature Pyramid Matching}
\newacronym{ds}{DS}{Discrepancy Scaling}
\newacronym{ad}{AD}{Anomaly Detection}
\newacronym{al}{AL}{Anomaly Localization}
\newacronym{asr}{ASR}{Automatic Speech Recognition}
\newacronym{fpm}{FPM}{Feature Pyramid Matching}
\newacronym{mse}{MSE}{Mean Squared Error}
\newacronym{poi}{POI}{Person-of-Interest}

\begin{document}
\bstctlcite{IEEEexample:BSTcontrol}

\title{Anomaly Detection and Localization for \\Speech Deepfakes via Feature Pyramid Matching\\

\thanks{
This work was supported by the FOSTERER project, funded by the Italian Ministry of Education, University, and Research within the PRIN 2022 program.
This work was partially supported by the European Union - Next Generation EU under the Italian National Recovery and Resilience Plan (NRRP), Mission 4, Component 2, Investment 1.3, CUP D43C22003080001, partnership on “Telecommunications of the Future” (PE00000001 - program “RESTART”).
This work was partially supported by the European Union - Next Generation EU under the Italian National Recovery and Resilience Plan (NRRP), Mission 4, Component 2, Investment 1.3, CUP D43C22003050001, partnership on ``SEcurity and RIghts in the CyberSpace’’ (PE00000014 - program ``FF4ALL-SERICS’’).
}
}

\author{\IEEEauthorblockN{
Emma Coletta,
Davide Salvi,
Viola Negroni,
Daniele Ugo Leonzio,
Paolo Bestagini
}
\IEEEauthorblockA{\textit{Dipartimento di Elettronica, Informazione e Bioingegneria (DEIB), Politecnico di Milano}\\
Piazza Leonardo Da Vinci 32, 20133 Milano, Italy\\
emma.coletta@mail.polimi.it, \{davide.salvi, viola.negroni, danieleugo.leonzio, paolo.bestagini\}@polimi.it
}
}

\maketitle

\begin{abstract}
The rise of AI-driven generative models has enabled the creation of highly realistic speech deepfakes—synthetic audio signals that can imitate target speakers' voices—raising critical security concerns.
Existing methods for detecting speech deepfakes primarily rely on supervised learning, which suffers from two critical limitations: limited generalization to unseen synthesis techniques and a lack of explainability.
In this paper, we address these issues by introducing a novel interpretable one-class detection framework, which reframes speech deepfake detection as an anomaly detection task.
Our model is trained exclusively on real speech to characterize its distribution, enabling the classification of out-of-distribution samples as synthetically generated.
Additionally, our framework produces interpretable anomaly maps during inference, highlighting anomalous regions across both time and frequency domains.
This is done through a Student-Teacher Feature Pyramid Matching system, enhanced with Discrepancy Scaling to improve generalization capabilities across unseen data distributions.
Extensive evaluations demonstrate the superior performance of our approach compared to the considered baselines, validating the effectiveness of framing speech deepfake detection as an anomaly detection problem.
\end{abstract}


\input{1_introduction}

\input{2_method}
\input{3_experimental_setup}

\input{4_results}
\input{5_conclusion}

\bibliographystyle{IEEEtran}
\bibliography{bstcontrol.bib, biblio.bib}

\end{document}

%% file: 1_introduction.tex
\section{Introduction}

In recent years, the rapid advancement of AI-driven generative systems has made the creation of high-quality synthetic content easier than ever.
While the democratization of generative tools brings numerous benefits, it also introduces serious social risks.
Malicious actors could exploit these technologies to produce deceptive synthetic media, enabling identity theft, financial fraud, and disinformation campaigns~\cite{amerini2025deepfake}.
Among these threats, speech deepfakes represent a particularly concerning issue in the audio domain. These are synthetically generated speech samples that mimic the voice of a target speaker, making them say arbitrary utterances.
Speech deepfakes have already been used in fraud, blackmail, and other security-related threats, prompting the multimedia forensics community to actively develop deepfake detection systems to preserve the integrity of digital communications~\cite{cuccovillo2022open}.

Diverse strategies have been proposed for speech deepfake detection, spanning from the analysis of low-level acoustic features~\cite{mari2023all, Sun_2023_CVPR} to that of higher-level semantic aspects~\cite{conti2022deepfake}.
These approaches leverage diverse frameworks, including traditional machine learning techniques~\cite{hamza2022deepfake}, advanced DL architectures~\cite{zaman2024hybrid, cuccovillo2024audio, negroni2024leveraging}, and pre-trained models~\cite{guo2024audio, salvi2024comparative}. 
Most existing detection methods follow a supervised learning strategy, where models are trained on labeled datasets containing both real and synthetic speech samples. 
However, this approach typically presents two critical challenges: generalization and interpretability~\cite{muller2022does}.
Generalization refers to the model's ability to detect deepfakes generated by speech synthesis methods not encountered during training. Interpretability concerns the ``black box'' nature of the detection systems, which typically provide little insight into the rationale underlying their predictions.
These two limitations significantly reduce the practical effectiveness of deepfake detectors in real-world scenarios.

Recent research has explored novel strategies to address these limitations.
One promising direction involves \gls{oc} classification systems to improve generalization \cite{kim2024one, lu2024one}. These models are trained exclusively on authentic speech data, learning its distribution so that any sample deviating from it is classified as synthetic.
On the other hand, efforts to enhance interpretability have focused on identifying the specific aspects of the audio signal that drive model predictions, improving the transparency of the detection process~\cite{salvi2023towards}.

In this paper, we propose a novel method that addresses these challenges through an interpretable, \gls{oc} speech deepfake detector.
Specifically, we reframe the detection problem as an \gls{ad} task, training a \gls{oc} model exclusively on real speech samples.
This enables the model to learn the statistical distribution of genuine speech and classify out-of-distribution samples as fake, preventing it from being fooled by unseen synthesis methods.
Our approach builds upon the \gls{stfpm} framework proposed in~\cite{valjakka2023anomaly}, enhanced with \gls{ds}~\cite{myllari2023discrepancy}, and is capable of detecting and localizing speech anomalies in both time and frequency domains.
To evaluate our approach, we test it across diverse scenarios, comparing its performance against multiple baselines.
Our results indicate its superiority and validate the effectiveness of framing speech deepfake detection as an \gls{ad} problem.

%% file: 2_method.tex
\section{Proposed Method}
\label{sec:method}

In this work, we address the problem of speech deepfake detection and propose an \gls{ad} framework to tackle it.
Our approach is based on the hypothesis that synthetic speech can be conceptualized as authentic speech embedded with anomalies, which arise from imperfections in the speech generative process.
These imperfections manifest as subtle artifacts that an analyst can leverage to discriminate between synthetic and authentic signals.
Traditional supervised learning systems explicitly learn the characteristics of these artifacts by training on labeled datasets containing both real and synthetic samples.
Then, at inference time, they use their acquired knowledge to differentiate between the two classes.
Although effective in controlled scenarios, this approach has significant limitations, particularly in its ability to generalize detection performance over speech generation methods that are unseen during training.
Different generative models introduce distinct artifacts, and detectors trained on a limited set of synthesis techniques may fail to recognize deepfakes produced by other methods.
To overcome this limitation, the proposed framework adopts a \gls{oc} learning paradigm, considering only authentic speech data during training.
The goal is to characterize the statistical distribution of genuine speech so that, during inference, any deviation from this distribution is flagged as an anomaly, indicating potential deepfake content.
By focusing only on authentic speech, the model's performance becomes independent of any speech synthesis method, enhancing its generalization capabilities.
Moreover, our \gls{ad} framework enhances interpretability by generating anomaly maps that localize these deviations in both time and frequency domains, providing valuable forensic insights into the input speech signal.

\subsection{Problem Formulation}
\label{subsec:problem}

The speech deepfake detection problem can be formally defined as follows. 
Let us consider a discrete-time speech signal $\mathbf{x}$ associated with a class $y \in \{0, 1\}$, where $0$ denotes that the signal is authentic and $1$ indicates that it has been synthetically generated.
The goal of this task is to develop a detector $\mathcal{D}$ capable of estimating the class of the signal $\mathbf{x}$ as $\hat{y}$,
% $\hat{y} \in [0,1]$, 
where $\hat{y}$ is the likelihood of the signal $\mathbf{x}$ being fake.

\glsreset{oc}
\glsreset{ad}
\glsreset{ds}

\subsection{Proposed System}
\label{subsec:system}

We address the speech deepfake detection task by adopting a \gls{oc} \gls{ad} strategy.
In this framework, the detector $\mathcal{D}$ is trained exclusively on authentic speech samples to learn their underlying statistical distribution.
During inference, the model evaluates how much a given speech signal deviates from this learned distribution.
Signals that closely align with the distribution of genuine speech are assigned a low $\hat{y}$ value, indicating authenticity. 
Conversely, signals exhibiting significant deviations are assigned a higher $\hat{y}$ value, suggesting the presence of deepfake content.

\begin{figure}
    \centering
    \includegraphics[width=0.95\columnwidth]{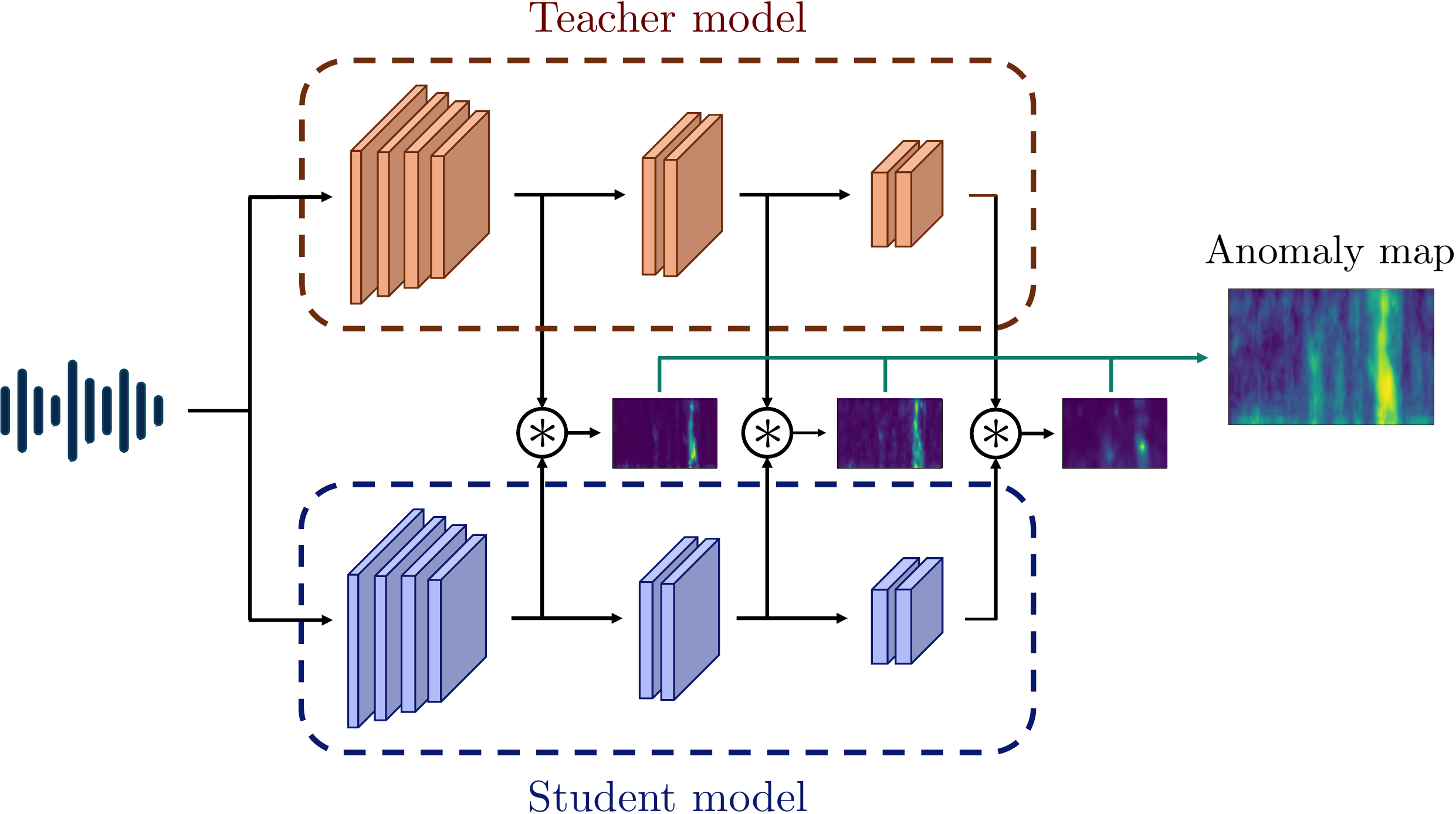}
    \caption{Pipeline of the proposed One-Class Anomaly Detection framework for speech deepfake detection.}
    \label{fig:pipeline}
    \vspace{-1em}
\end{figure}

To implement the proposed framework, we adopt a Student-Teacher architecture inspired by the \gls{stfpm} method introduced in~\cite{wang2021student_teacher}.
As illustrated in Figure~\ref{fig:pipeline}, the system consists of two identical CNN-based networks, i.e., the teacher and the student, which process acoustic features extracted from an input speech signal.
The teacher network is pre-trained on the speaker identification task, enabling it to learn robust, speech-specific representations.
The student network is then trained to replicate the teacher's activations through \gls{fpm}~\cite{valjakka2023anomaly}.
This training process encourages the student to mimic the teacher’s feature representations. 
However, since the two models are trained exclusively on authentic data, the student learns to approximate the teacher’s behavior only for real speech samples. 
During inference, when presented with synthetic speech, the student will struggle to replicate the teacher’s feature maps, resulting in discrepancies between their activations.
These discrepancies are quantified as anomaly scores, which we use to detect deepfakes.

To compute these scores, we leverage \gls{fpm}, which compares the activations of the teacher and student networks at multiple designated layers.
This multi-level comparison captures anomalies at different scales: lower layers encode fine-grained, low-level acoustic details, while higher layers capture broader, context-rich information.
The discrepancy between the normalized activations of the teacher and student models serves as the anomaly indicator.
Ideally, the anomaly values will be low when the system receives real speech, and high when the input is synthetic.
Finally, these discrepancies can be visualized through an anomaly map, highlighting time-frequency regions where generation artifacts are most prominent.
Regions with higher anomaly values indicate areas in the signal where synthetic speech deviates most significantly from real speech, providing an interpretable way to analyze deepfake detection.

To further enhance the performance of our \gls{ad} model, we incorporate \gls{ds}~\cite{myllari2023discrepancy}.
This technique standardizes the discrepancy values between the teacher and student networks during inference based on statistical properties of the test data.
The normalization process involves computing the mean and standard deviation of the test distribution using a small calibration set.
\gls{ds} is applied layer by layer to both the teacher and student networks, amplifying subtle anomalies and enabling the detection of fine-grained synthetic artifacts.
In our analysis, we use \gls{ds} as a calibration strategy to enhance detection performances on out-of-distribution data, as shown in our experimental results in Section~\ref{sec:results}.

%% file: 3_experimental_setup.tex
\glsreset{ds}

\section{Experimental Setup}
\label{sec:setup}

In this section, we describe the evaluation setup used in our experiments.
We begin by detailing the architecture and the strategies used to train and evaluate the teacher and student networks.
Then, we introduce the baseline systems used to validate our findings.
Finally, we present the datasets we employed in our experimental campaign. 

\subsection{Training and Evaluation Strategies}

The model that we considered for both teacher and student networks is a modified version of ResNet18~\cite{he2016deep}, adapted to process 2D acoustic features.
It incorporates a dropout layer for regularization, a fully connected layer with \num{256} units, and a ReLU activation function.
We considered mel-spectrograms as the acoustic feature representation, as they closely mimic human auditory perception.
This choice enhances the interpretability of the generated anomaly maps, making them more reflective of how anomalies might be perceived by listeners.
The networks are trained and tested on audio segments of \SI{4}{\second}.

We trained the teacher network on the speaker identification task using the LibriSpeech dataset~\cite{panayotov2015librispeech}.
We utilize the \textit{train-clean-360} subset, which contains speech from \num{921} speakers.
This strategy is inspired by~\cite{valjakka2023anomaly}, where the backbone was pre-trained on the image classification task. 
In our implementation, we decided to consider a pre-training that better aligns with the speech domain.
The network has been trained for \num{300} epochs with an early stopping of \num{15} epochs, considering a batch size of \num{64} samples and a learning rate of $10^{-4}$. We assumed Cross Entropy as the loss function, with AdamW as optimizer and Cosine Annealing as learning rate scheduler.

The student network shares the same architecture of the teacher model, except for the final classification layer, which is removed since it is not needed for \gls{ad}.
It is trained to minimize the discrepancy between its feature maps and those of the teacher network at the three final convolutional layers.
This is done using a feature-matching loss, which computes the \gls{mse} between the L2-normalized activations of the teacher and student models.
During training, both networks process the same input and generate their respective feature representations.
The weights of the teacher remain frozen, while those of the student are updated.
The loss value is computed by averaging the squared differences across feature maps.
All the other hyperparameters are identical to those used for training the teacher.
At inference time, the discrepancy between the teacher and student activations is used to generate an anomaly map.
The average value of this map serves as the anomaly score for binary classification.

To enhance the anomaly detection capabilities of the proposed framework, we integrate \gls{ds}.
\gls{ds} computes the mean and standard deviation of the discrepancies between the teacher and student activations on a small calibration dataset of real speech tracks.
Then, during inference, these statistics are used to normalize the anomaly scores at each layer, improving the sensitivity to synthetic artifacts and enhancing the model's generalization capabilities on out-of-distribution data.

\subsection{Baseline systems}

To validate the effectiveness of our proposed framework, we compare its performance against two baselines.
The first one is RawNet2~\cite{tak2021end}, a widely used supervised learning model designed to differentiate between real and synthetic speech samples.
The second baseline is the same ResNet18 model considered above, trained following the \gls{oc} approach introduced in~\cite{kim2024one}.
Unlike the original implementation, we train this model using a strict \gls{oc} setup, excluding synthetic samples during training. This ensures a fair comparison with our proposed method, which also relies exclusively on real data.
These baselines were selected to highlight two key aspects: the improved generalization capability of a \gls{oc} model compared to a traditional supervised classifier, and the advantages of the proposed \gls{ad} framework over an existing \gls{oc} approach.

\subsection{Datasets}
\label{subsec:dataset}

To assess the generalization capabilities of the considered systems, we conducted experiments across multiple speech deepfake datasets, with all audio sampled at \SI{16}{\kilo\hertz}.

We performed two distinct types of analyses: single and multi-speaker experiments.
In single-speaker experiments, we focus on a specific target voice for both training and testing the detectors.
This scenario simulates a \gls{poi} use case~\cite{agarwal2019protecting}, where the goal is to protect a specific individual from deepfake attacks by leveraging their unique voice characteristics.
For this analysis, we considered the voice of Linda Johnson from LJSpeech~\cite{ljspeech17} as real data source.
The synthetic datasets used for evaluation include TIMIT-TTS~\cite{salvi2023timit}, Purdue speech dataset~\cite{bhagtani2024recent}, and MLAAD~\cite{muller2024mlaad}, focusing on their subsets which contain deepfake speech generated to mimic the target voice.
In this scenario, the \gls{oc} detectors were trained on a partition of LJSpeech excluded from the evaluation, while RawNet2 was trained on the same LJSpeech partition combined with a subset of TIMIT-TTS.

In multi-speaker experiments, we tested generalization across diverse voices using ASVspoof 2019~\cite{wang2020asvspoof}, In-the-Wild~\cite{muller2022does}, FakeOrReal~\cite{reimao2019dataset}, and the complete Purdue speech dataset~\cite{bhagtani2024recent}.
In this setting, models were trained on the \textit{train} and \textit{dev} partitions of ASVspoof 2019. For \gls{oc} detectors, only authentic speech data was used during training, while RawNet2 was trained on both real and fake samples.

%% file: 4_results.tex
\section{Results}
\label{sec:results}

In this section, we analyze and discuss the performance of our proposed framework for speech deepfake detection, using \gls{roc} curves, \gls{auc}, and \gls{eer}.
Additionally, we assess the interpretability of our approach by examining the anomaly maps it produces.

\begin{figure}
    \centering
    \includegraphics[width=0.6\columnwidth]{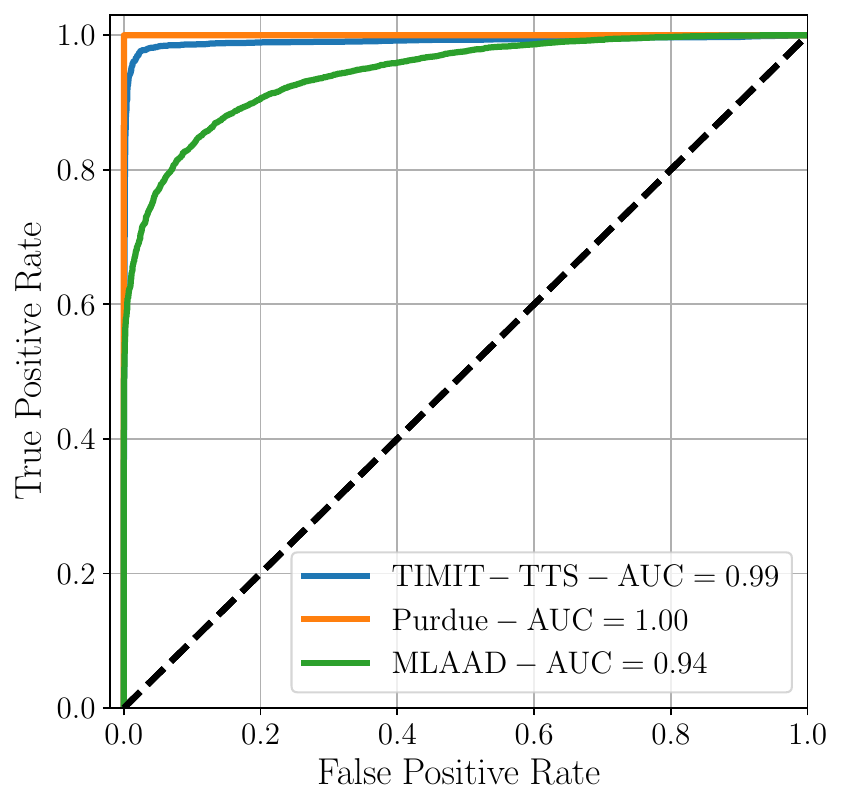}
    \caption{\gls{roc} curves of the proposed framework without DS evaluated in a single-speaker scenario.}
    \label{fig:single_sp_roc}
    \vspace{-0.5em}
\end{figure}

\begin{figure}
    \centering
    \includegraphics[width=0.9\columnwidth]{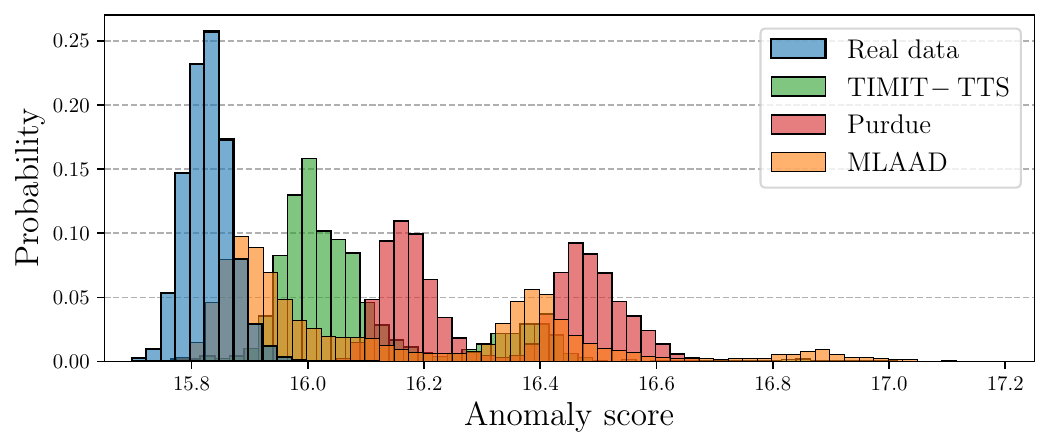}
    \caption{Anomaly score distributions of the proposed framework without DS evaluated in a single-speaker scenario.}
    \label{fig:single_sp_distrib}
\end{figure}

\vspace{0.2em}\noindent \textbf{Single-Speaker Scenario.}
As a first experiment, we evaluate the system in a single-speaker scenario, simulating a \gls{poi} use case.
We do this as an initial analysis because we hypothesize that characterizing the voice of a single target speaker may be easier for a \gls{oc} model, compared to modeling multiple voices simultaneously.
Table~\ref{tab:detection_single_sp} shows the complete results of this analysis.
Our proposed framework, both with and without \gls{ds}, outperforms the baselines across most datasets.
On the TIMIT-TTS dataset, RawNet2 achieves the best results, which is expected given its supervised training on in-domain data. However, the model performance degrades significantly on the other datasets.
On the other hand, the \gls{oc} baseline shows reasonable performance on MLAAD, but performs the worst overall, demonstrating its limitations in a \gls{poi} scenario.
Our proposed framework without \gls{ds} achieves the best performance, with perfect detection on Purdue and strong results on both TIMIT-TTS and MLAAD.
The \gls{roc} curves and anomaly score distributions for this configuration are shown in Fig.~\ref{fig:single_sp_roc} and Fig.~\ref{fig:single_sp_distrib}, respectively.
In this scenario, applying \gls{ds} slightly reduces performance on some datasets but still maintains high detection accuracy.

\begin{table}
\centering
\caption{Single-speaker detection performance (\%). Best values are bold, second-best are italic.}
\label{tab:detection_single_sp}
\resizebox{.9\columnwidth}{!}{
\begin{tabular}{lcccccc}
\toprule
         & \multicolumn{2}{c}{TIMIT-TTS}     & \multicolumn{2}{c}{Purdue}       & \multicolumn{2}{c}{MLAAD}      \\ 
\cmidrule(lr){2-3} \cmidrule(lr){4-5} \cmidrule(lr){6-7} 
         & AUC $\uparrow$ & EER $\downarrow$ & AUC $\uparrow$ & EER $\downarrow$ & AUC $\uparrow$ & EER $\downarrow$ \\ \midrule
RawNet2~\cite{tak2021end}    & \textbf{99.9}  & \textbf{0.5}   & 70.6  & 35.1     & 62.2  & 43.4   \\
OC baseline~\cite{kim2024one}   & 56.8  & 46.9   & 62.4  & 39.8     & 78.1  & 29.7  \\
\textbf{Ours (w/o DS)}          & \emph{99.1}  & \emph{2.4}   & \textbf{100.0}  & \textbf{0.0}     & \textbf{94.3}  & \textbf{13.3}    \\
\textbf{Ours (with DS)}         & 96.3  & 9.7   & \textbf{100.0}  & \emph{0.1}     & \emph{86.0}  & \emph{23.6}    \\ 
\bottomrule
\end{tabular}}
\end{table}

\begin{figure}
    \centering
    \includegraphics[width=0.6\columnwidth]{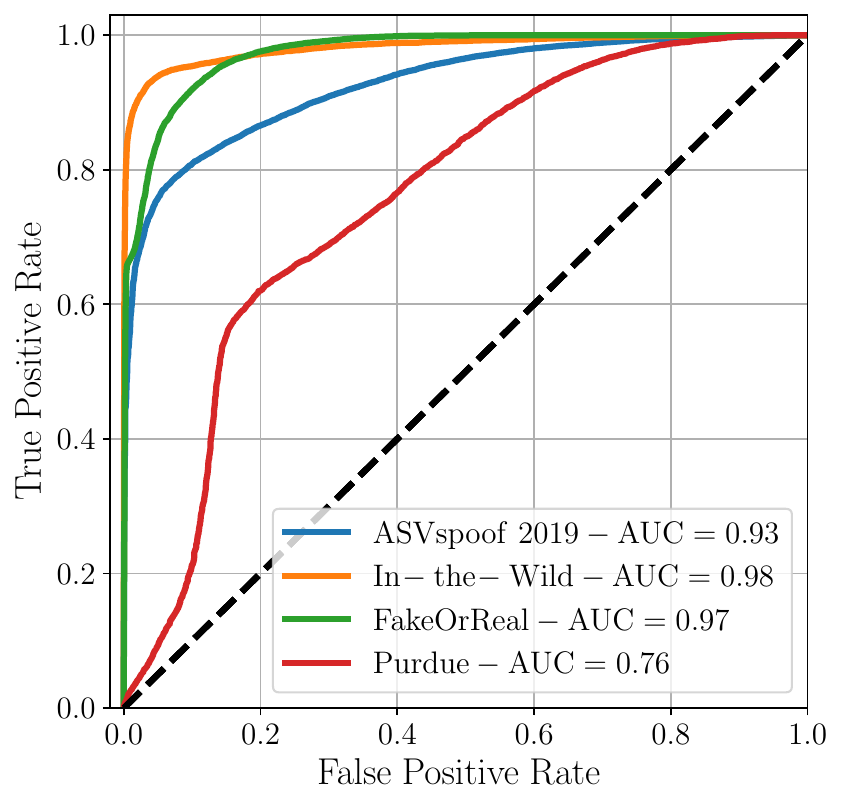}
    \caption{\gls{roc} curves of the proposed framework with \gls{ds} evaluated in a multi-speaker scenario.}
    \label{fig:multi_sp_roc}
    \vspace{-0.5em}
\end{figure}

\vspace{0.2em}\noindent \textbf{Multi-Speaker Scenario.}
Given the strong performance in the single-speaker setting, we extend our evaluation to a multi-speaker scenario, where we expect that detecting deepfakes becomes more challenging due to the increased variability in real speech.
Table~\ref{tab:detection_multi_sp} presents the results of this analysis.
Again, RawNet2 performs best on its training dataset (ASVspoof 2019) but struggles on unseen datasets, confirming the limitations of supervised learning methods.
The \gls{oc} baseline performs better than in the single-speaker case, especially on the FakeOrReal dataset. This suggests that the model is more effective at characterizing multiple voices together than it is in a \gls{poi} context.
% The \gls{oc} baseline performs better than in the single-speaker case, particularly on FakeOrReal, indicating that characterizing multiple voices together can improve detection compared to focusing on a single target voice.
The best-performing model is our proposed framework with \gls{ds}, achieving the best detection results on all datasets except ASVspoof 2019. The \gls{roc} curves for this model are shown in Figure~\ref{fig:multi_sp_roc}.
These results demonstrate the effectiveness of \gls{ds} in enhancing the model’s ability to detect deepfakes, particularly in datasets where the real speech distribution differs significantly from the training data.

\begin{table}
\centering
\caption{Multi-speaker detection performance (\%). Best values are bold, second-best are italic.}
\label{tab:detection_multi_sp}
\resizebox{\columnwidth}{!}{
\begin{tabular}{lcccccccc}
\toprule
         & \multicolumn{2}{c}{ASVspoof 2019}     & \multicolumn{2}{c}{In-the-Wild}       & \multicolumn{2}{c}{FakeOrReal}       & \multicolumn{2}{c}{Purdue}  \\ \cmidrule(lr){2-3} \cmidrule(lr){4-5} \cmidrule(lr){6-7} \cmidrule(lr){8-9}
         & AUC $\uparrow$ & EER $\downarrow$ & AUC $\uparrow$ & EER $\downarrow$ & AUC $\uparrow$ & EER $\downarrow$ & AUC $\uparrow$ & EER $\downarrow$ \\ \midrule
        RawNet2~\cite{tak2021end}        & \textbf{93.2}  & \textbf{10.6}   & 63.9  & 40.9   & 82.1   & 25.1   & 56.3    & 44.9 \\
        OC baseline~\cite{kim2024one}        & 91.9  & \emph{14.8}   & 79.0  & 28.9   & \textbf{99.0}   & \textbf{5.5}   & \emph{66.6}    & \emph{35.9} \\
        \textbf{Ours (w/o DS)}      & 88.1  & 21.2   & \emph{95.1}  & \emph{11.3}   & 92.8   & 18.0   & 63.8    & 38.2          \\
        \textbf{Ours (with  DS)}    & \emph{92.7}  & 15.6   & \textbf{98.2}  & \textbf{5.7}   & \emph{97.4}   & \emph{9.1}   & \textbf{75.7}    & \textbf{30.6} \\ \bottomrule
\end{tabular}}
\end{table}

% OLD RESULTS
% \begin{table}
% \centering
% \caption{Detection performance of the considered systems in a multi-speaker scenario (\%).}
% \label{tab:detection_multi_sp}
% \resizebox{\columnwidth}{!}{
% \begin{tabular}{lcccccccc}
% \toprule
%          & \multicolumn{2}{c}{ASVspoof 2019}     & \multicolumn{2}{c}{In-the-Wild}       & \multicolumn{2}{c}{FakeOrReal}       & \multicolumn{2}{c}{Purdue}  \\ \cmidrule(lr){2-3} \cmidrule(lr){4-5} \cmidrule(lr){6-7} \cmidrule(lr){8-9}
%          & AUC $\uparrow$ & EER $\downarrow$ & AUC $\uparrow$ & EER $\downarrow$ & AUC $\uparrow$ & EER $\downarrow$ & AUC $\uparrow$ & EER $\downarrow$ \\ \midrule
%         RawNet2~\cite{tak2021end}        & 91.9  & 10.6   & 63.9  & 40.9   & 82.1   & 25.1   & 56.3    & 44.9 \\
%         OC baseline~\cite{kim2024one}        & \textbf{93.2}  & \textbf{14.8}   & 79.0  & 28.9   & \textbf{99.0}   & \textbf{5.5}   & 66.6    & 35.9 \\
%         \textbf{Ours (w/o DS)}      & 87.1  & 22.2   & 95.1  & 11.6   & 92.8   & 18.0   & 61.8    & 40.2          \\
%         \textbf{Ours (with  DS)}    & 92.7  & 15.6   & \textbf{98.2}  & \textbf{5.7}   & 97.4   & 9.1   & \textbf{75.7}    & \textbf{30.6} \\ \bottomrule
% \end{tabular}}
% \end{table}

\begin{figure*}
    \centering
    % First Subfigure
    \begin{subfigure}{0.24\textwidth}
        \includegraphics[width=\linewidth]{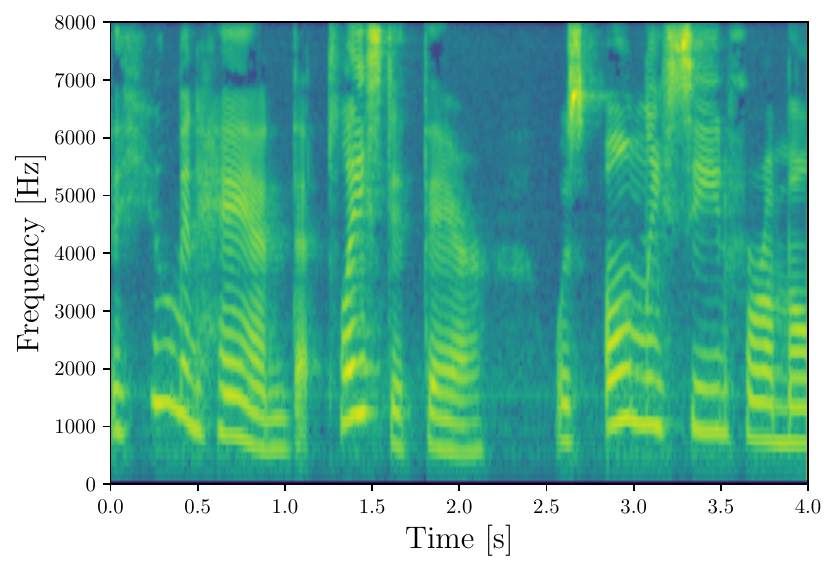}
        \caption{Real audio Mel-spectrogram}
    \end{subfigure}
    % Second Subfigure
    \begin{subfigure}{0.24\textwidth}
        \includegraphics[width=\linewidth]{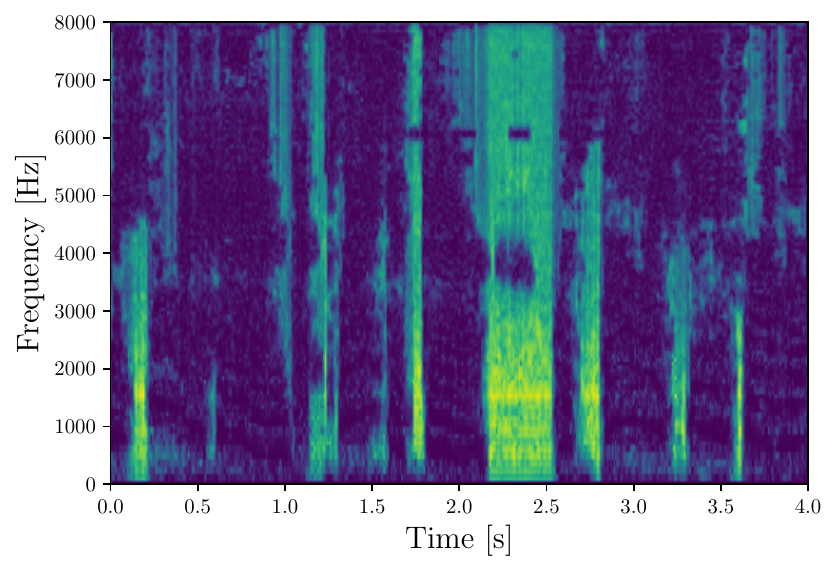}
        \caption{Ground truth}
    \end{subfigure}
    % Third Subfigure
    \begin{subfigure}{0.24\textwidth}
        \includegraphics[width=\linewidth]{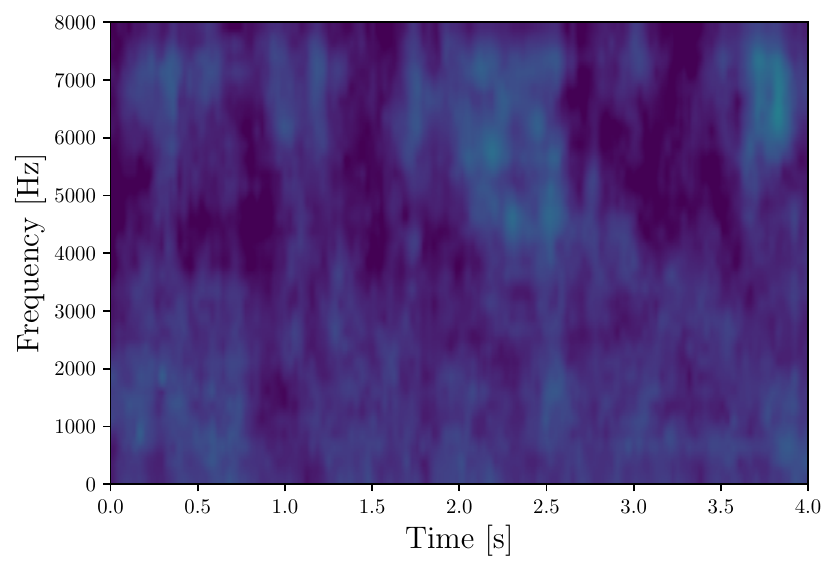}
        \caption{Real audio anomaly map}
    \end{subfigure}
    % Fourth Subfigure
    \begin{subfigure}{0.24\textwidth}
        \includegraphics[width=\linewidth]{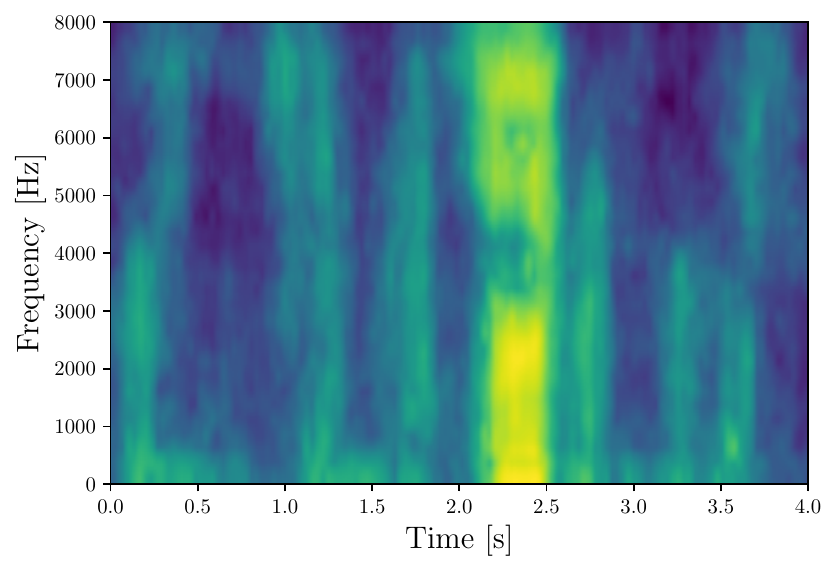}
        \caption{Fake audio anomaly map}
    \end{subfigure}
    
    \caption{Anomaly localization results for a real audio track and its fake counterpart.}
    \label{fig:anomaly_map}
    %\vspace{-.5em}
\end{figure*}

\vspace{0.2em}\noindent \textbf{Interpretability Analysis.}
As a final experiment, we assess the interpretability of our framework by analyzing the anomaly maps it produces.
To do this, we take a real audio track from LJSpeech, compute its mel-spectrogram, and re-synthesize it using the Waveglow~\cite{prenger2019waveglow} vocoder.
Following the approach in~\cite{wang2023spoofed}, we consider this re-synthesized track as synthetic, even though it is derived from real speech features.
This method ensures that the real and fake tracks are perfectly time-aligned, allowing us to compute their difference and generate a ground truth anomaly map.
We then use our framework to estimate the predicted anomaly map.
Figure \ref{fig:anomaly_map} shows the results of this analysis, including the mel-spectrogram of the original audio, the ground truth anomaly map, and the anomaly maps produced by the system for both the real and fake tracks.
The anomaly map for the authentic track is nearly empty, indicating no detected anomalies, while the map for the fake track closely matches the ground truth, highlighting the system's ability to localize synthetic artifacts.

%% file: 5_conclusion.tex
\glsreset{oc}
\glsreset{ad}
\glsreset{ds}

\section{Conclusions}
\label{sec:conclusion}

In this work, we proposed a novel \gls{oc} \gls{ad} framework for speech deepfake detection, addressing the generalization and interpretability limitations of traditional supervised approaches.
Our system, based on the \gls{stfpm} architecture, is trained exclusively on authentic speech to characterize its statistical distribution and detect anomalies attributable to synthetic content.
Experimental results show that our method outperforms both supervised classifiers (RawNet2) and an alternative \gls{oc} model.
By incorporating \gls{ds}, we further enhance the system's generalization capabilities on out-of-distribution data.
Additionally, the framework generates anomaly maps that localize artifacts in time and frequency, improving the interpretability of the results.
Our findings validate the effectiveness of framing speech deepfake detection as an \gls{ad} problem, offering superior generalization over traditional supervised methods. Future work will refine \gls{ds} and explore applications to diverse real-world forensic scenarios.

%% file: 0_main.bbl
\begin{thebibliography}{10}
\providecommand{\url}[1]{#1}
\csname url@samestyle\endcsname
\providecommand{\newblock}{\relax}
\providecommand{\bibinfo}[2]{#2}
\providecommand{\BIBentrySTDinterwordspacing}{\spaceskip=0pt\relax}
\providecommand{\BIBentryALTinterwordstretchfactor}{4}
\providecommand{\BIBentryALTinterwordspacing}{\spaceskip=\fontdimen2\font plus
\BIBentryALTinterwordstretchfactor\fontdimen3\font minus \fontdimen4\font\relax}
\providecommand{\BIBforeignlanguage}[2]{{%
\expandafter\ifx\csname l@#1\endcsname\relax
\typeout{** WARNING: IEEEtran.bst: No hyphenation pattern has been}%
\typeout{** loaded for the language `#1'. Using the pattern for}%
\typeout{** the default language instead.}%
\else
\language=\csname l@#1\endcsname
\fi
#2}}
\providecommand{\BIBdecl}{\relax}
\BIBdecl

\bibitem{amerini2025deepfake}
I.~Amerini, M.~Barni, S.~Battiato \emph{et~al.}, ``Deepfake media forensics: Status and future challenges,'' \emph{Journal of Imaging}, 2025.

\bibitem{cuccovillo2022open}
L.~Cuccovillo, C.~Papastergiopoulos, A.~Vafeiadis \emph{et~al.}, ``Open challenges in synthetic speech detection,'' in \emph{IEEE International Workshop on Information Forensics and Security (WIFS)}, 2022.

\bibitem{mari2023all}
D.~Mari, D.~Salvi, P.~Bestagini, and S.~Milani, ``{All-for-one and one-for-all: Deep learning-based feature fusion for synthetic speech detection},'' in \emph{Joint European Conference on Machine Learning and Knowledge Discovery in Databases (ECML PKDD)}, 2023.

\bibitem{Sun_2023_CVPR}
C.~Sun, S.~Jia, S.~Hou, and S.~Lyu, ``{AI-Synthesized Voice Detection Using Neural Vocoder Artifacts},'' in \emph{IEEE/CVF Conference on Computer Vision and Pattern Recognition Workshop (CVPRW)}, 2023.

\bibitem{conti2022deepfake}
E.~Conti, D.~Salvi, C.~Borrelli \emph{et~al.}, ``Deepfake speech detection through emotion recognition: a semantic approach,'' in \emph{IEEE International Conference on Acoustics, Speech and Signal Processing (ICASSP)}, 2022.

\bibitem{hamza2022deepfake}
A.~Hamza, A.~Javed, F.~Iqbal \emph{et~al.}, ``{Deepfake audio detection via MFCC features using machine learning},'' \emph{IEEE Access}, 2022.

\bibitem{zaman2024hybrid}
K.~Zaman, I.~J. Samiul, M.~Sah \emph{et~al.}, ``Hybrid transformer architectures with diverse audio features for deepfake speech classification,'' \emph{IEEE Access}, 2024.

\bibitem{cuccovillo2024audio}
L.~Cuccovillo, M.~Gerhardt, and P.~Aichroth, ``Audio transformer for synthetic speech detection via multi-formant analysis,'' in \emph{IEEE/CVF Conference on Computer Vision and Pattern Recognition Workshops (CVPRW)}, 2024.

\bibitem{negroni2024leveraging}
V.~Negroni, D.~Salvi, A.~I. Mezza \emph{et~al.}, ``Leveraging mixture of experts for improved speech deepfake detection,'' in \emph{IEEE International Conference on Acoustics, Speech and Signal Processing (ICASSP)}, 2025.

\bibitem{guo2024audio}
Y.~Guo, H.~Huang, X.~Chen \emph{et~al.}, ``{Audio Deepfake Detection With Self-Supervised Wavlm And Multi-Fusion Attentive Classifier},'' in \emph{IEEE International Conference on Acoustics, Speech and Signal Processing (ICASSP)}, 2024.

\bibitem{salvi2024comparative}
D.~Salvi, A.~K.~S. Yadav, K.~Bhagtani \emph{et~al.}, ``{Comparative Analysis of ASR Methods for Speech Deepfake Detection},'' in \emph{Asilomar Conference on Signals, Systems, and Computers}, 2024.

\bibitem{muller2022does}
N.~M. M{\"u}ller, P.~Czempin, F.~Dieckmann \emph{et~al.}, ``Does audio deepfake detection generalize?'' in \emph{Interspeech}, 2022.

\bibitem{kim2024one}
H.~M. Kim, K.~Jang, and H.~Kim, ``{One-class learning with adaptive centroid shift for audio deepfake detection},'' \emph{Interspeech}, 2024.

\bibitem{lu2024one}
J.~Lu, Y.~Zhang, W.~Wang \emph{et~al.}, ``{One-Class Knowledge Distillation for Spoofing Speech Detection},'' in \emph{IEEE International Conference on Acoustics, Speech and Signal Processing (ICASSP)}, 2024.

\bibitem{salvi2023towards}
D.~Salvi, P.~Bestagini, and S.~Tubaro, ``Towards frequency band explainability in synthetic speech detection,'' in \emph{European Signal Processing Conference (EUSIPCO)}.\hskip 1em plus 0.5em minus 0.4em\relax IEEE, 2023.

\bibitem{valjakka2023anomaly}
J.~Valjakka, J.~Myll{\"a}ri, L.~Myllyaho \emph{et~al.}, ``Anomaly localization in audio via feature pyramid matching,'' in \emph{IEEE Annual Computers, Software, and Applications Conference}, 2023.

\bibitem{myllari2023discrepancy}
J.~Myll{\"a}ri and J.~K. Nurminen, ``Discrepancy scaling for fast unsupervised anomaly localization,'' in \emph{IEEE Annual Computers, Software, and Applications Conference}, 2023.

\bibitem{wang2021student_teacher}
G.~Wang, S.~Han, E.~Ding, and D.~Huang, ``Student-teacher feature pyramid matching for anomaly detection,'' in \emph{The British Machine Vision Conference (BMVC)}, 2021.

\bibitem{he2016deep}
K.~He, X.~Zhang, S.~Ren, and J.~Sun, ``Deep residual learning for image recognition,'' in \emph{IEEE Conference on Computer Vision and Pattern Recognition (CVPR)}, 2016.

\bibitem{panayotov2015librispeech}
V.~Panayotov, G.~Chen, D.~Povey, and S.~Khudanpur, ``Librispeech: an asr corpus based on public domain audio books,'' in \emph{IEEE International Conference on Acoustics, Speech and Signal Processing (ICASSP)}, 2015.

\bibitem{tak2021end}
H.~Tak, J.~Patino, M.~Todisco \emph{et~al.}, ``{End-to-end anti-spoofing with RawNet2},'' in \emph{IEEE International Conference on Acoustics, Speech and Signal Processing (ICASSP)}, 2021.

\bibitem{agarwal2019protecting}
S.~Agarwal, H.~Farid, Y.~Gu \emph{et~al.}, ``Protecting world leaders against deep fakes.'' in \emph{IEEE/CVF Conference on Computer Vision and Pattern Recognition Workshop (CVPRW)}, 2019.

\bibitem{ljspeech17}
K.~Ito and L.~Johnson, ``The lj speech dataset,'' \url{https://keithito.com/LJ-Speech-Dataset/}, 2017.

\bibitem{salvi2023timit}
D.~Salvi, B.~Hosler, P.~Bestagini \emph{et~al.}, ``{TIMIT-TTS: a Text-to-Speech Dataset for Multimodal Synthetic Media Detection},'' \emph{IEEE Access}, 2023.

\bibitem{bhagtani2024recent}
K.~Bhagtani, A.~K.~S. Yadav, P.~Bestagini, and E.~J. Delp, ``{Are Recent Deepfake Speech Generators Detectable?}'' in \emph{ACM Workshop on Information Hiding and Multimedia Security}, 2024.

\bibitem{muller2024mlaad}
N.~M. M{\"u}ller, P.~Kawa, W.~H. Choong \emph{et~al.}, ``{MLAAD: The Multi-Language Audio Anti-Spoofing Dataset},'' \emph{IEEE International Joint Conference on Neural Networks (IJCNN)}, 2024.

\bibitem{wang2020asvspoof}
X.~Wang, J.~Yamagishi, M.~Todisco \emph{et~al.}, ``Asvspoof 2019: A large-scale public database of synthesized, converted and replayed speech,'' \emph{Computer Speech \& Language}, vol.~64, p. 101114, 2020.

\bibitem{reimao2019dataset}
R.~Reimao and V.~Tzerpos, ``{FOR: A dataset for synthetic speech detection},'' in \emph{IEEE International Conference on Speech Technology and Human-Computer Dialogue (SpeD)}, 2019.

\bibitem{prenger2019waveglow}
R.~Prenger, R.~Valle, and B.~Catanzaro, ``{Waveglow: A flow-based generative network for speech synthesis},'' in \emph{IEEE International Conference on Acoustics, Speech and Signal Processing (ICASSP)}, 2019.

\bibitem{wang2023spoofed}
X.~Wang and J.~Yamagishi, ``Spoofed training data for speech spoofing countermeasure can be efficiently created using neural vocoders,'' in \emph{IEEE International Conference on Acoustics, Speech and Signal Processing (ICASSP)}, 2023.

\end{thebibliography}
